\documentclass[11pt]{article}
\pdfoutput=1
\usepackage{cite}
\usepackage{graphicx}
\usepackage{multicol}
\usepackage{amsfonts}
\usepackage{amssymb}
\usepackage{array}
\usepackage{amsmath}
\usepackage{heck}
\usepackage{hyperref}
\usepackage{setspace}
\usepackage{verbatim}
\usepackage{color}
\usepackage{float}
\usepackage{epsfig}

\newcommand{\ba}{\begin{eqnarray}}
\newcommand{\ea}{\end{eqnarray}}

\begin{document}

\def\simgt{\mathrel{\lower2.5pt\vbox{\lineskip=0pt\baselineskip=0pt
           \hbox{$>$}\hbox{$\sim$}}}}
\def\simlt{\mathrel{\lower2.5pt\vbox{\lineskip=0pt\baselineskip=0pt
           \hbox{$<$}\hbox{$\sim$}}}}

\begin{titlepage}

\begin{flushright}
\end{flushright}

\begin{center}

{\LARGE \bf R-parity Conservation from a Top Down Perspective}

\vskip 1cm

{\large Bobby S. Acharya$^{\dag}$, Gordon L. Kane$^{\star}$, Piyush Kumar$^{\S}$, Ran Lu$^{\ddagger}$, Bob Zheng$^{\star}$}

\vskip 0.4cm

$^{\dag}${\it Department of Physics, King's College London, London, UK and \\
International Centre for Theoretical Physics, Trieste, Italy}

\vskip 0.4cm

$^{*}${\it Department of Physics $\&$ Michigan Center for Theoretical Physics\\ University of Michigan, Ann Arbor \\
Ann Arbor, MI 48109 USA}

\vskip 0.4cm

$^{\S}${\it Department of Physics \\
Yale University, New Haven, CT 06520 USA}

\vskip 0.4cm

$^{\ddagger}${\it Department of Physics \\
University of Wisconsin, Madison, WI 53706 USA}

\abstract{Motivated by results from the LHC and dark matter searches, we study the possibility of phenomenologically viable R-parity violation in $SU(5)$ GUT models from a top-down point of view. We show that in contrast to the more model dependent bounds on the proton lifetime, the limits on neutrino masses provide a robust, stringent and complementary constraint on all $SU(5)$ GUT-based R-parity violating models. Focusing on well-motivated string/$M$ theory GUT frameworks with mechanisms for doublet-triplet splitting and a solution to the $\mu/B\mu$ problems, we show that imposing the neutrino mass bounds implies that R-parity violation is  disfavored. The arguments can also be generalized to minimal $SO(10)$ GUTs. An experimental observation of R-parity violation would, therefore, disfavor such classes of top-down GUT models.}

\end{center}
\maketitle

\end{titlepage}
\tableofcontents

\section{Introduction and Motivation}

The Large Hadron Collider (LHC) experiments as well as direct and indirect dark matter searches have set strong constraints on many classes of models for physics beyond the Standard Model (BSM). The lack of BSM signals in such experiments motivates us to re-assess our assumptions about the nature of dark matter and its interactions with Standard Model particles. In particular, in many models in which dark matter has electroweak interactions (i.e. is a WIMP), the stability of the dark matter particle is ensured by the presence of a discrete symmetry, for example, $R$-parity in the minimal supersymmetric Standard Model (MSSM). WIMP dark matter stabilised by such symmetries then would lead to a ``missing energy" signal at the LHC. From a bottom up point of view investigating the violation of such symmetries can be done ``operator by operator", but usually there are a large number of effective operators which break the symmetry, leading to significant model dependence. On the other hand, top down models can often provide boundary conditions arising from e.g. constraints from grand unification and/or the nature of supersymmetry breaking which can greatly reduce such model dependence.

In the Standard Model (SM), baryon number $(B)$ and lepton number $(L)$ violating processes are forbidden by accidental global symmetries of the Lagrangian. However, in the Minimal Supersymmetric Standard Model (MSSM) there is no such symmetry which forbids dangerous $B-L$ violating superpotential operators at the renormalizable level. Typically, an additional global symmetry commonly referred to as R-parity\cite{Salam:1974xa,Farrar:1978xj} is imposed to forbid the $B-L$ violating operators which might otherwise induce unacceptable proton decay \cite{Hinchliffe:1992ad}. In addition, R-parity conservation enforces the stability of the lightest supersymmetric particle (LSP), which allows the LSP to be a natural candidate for dark matter\cite{Goldberg:1983nd,Ellis:1983ew}. 

Despite these nice features, an exact R-parity is not a phenomenological necessity, as it is possible to construct (at least from the $\sim$TeV-scale point of view) R-parity violating (RPV) models naturally satisfying all phenomenological constraints. RPV supersymmetry also has potentially interesting phenomenological implications which distinguish it from the R-parity conserving alternative\cite{Aulakh:1982yn,Hall:1983id,Ellis:1984gi,Ross:1984yg,Dreiner:1997uz,Barbier:2004ez}. It is therefore pertinent to ask what theoretical structures are responsible for explaining the presence or absence of RPV terms in realistic SUSY models such as the MSSM.\footnote{For example, the recent work of \cite{Csaki:2011ge,Csaki:2013we} has proposed that the theoretical structure which determines R-parity violation in the MSSM has the same origin as the flavor structure of the R-parity conserving Yukawa couplings.} Given a particular theoretical framework, is it possible to use low energy constraints to obtain a correlated prediction between realistic models and the presence or absence of R-parity violation?

In this work, we study this question in Supersymmetric Grand Unified Theories (SUSY GUT's)\cite{Dimopoulos:1981yj}. Such a framework is well motivated by the apparent unification of gauge couplings in the MSSM, as well as many stringy/$M$-theoretic UV completions (see for example \cite{Raby:2011jt} and references therein). It also has the advantage of providing substantial theoretical constraints compared to a purely bottom-up study, due to the enhanced GUT gauge symmetries. Many GUT models suffer from phenomenological problems such as higher-dimensional proton-decay, doublet-triplet splitting, etc. to name a few. We will focus on classes of top-down GUT models in which natural mechanisms are available to address these phenomenological problems. In particular, we focus on $SU(5)$ GUT models which have i) a mechanism for breaking the GUT symmetry to that of the SM, ii) a mechanism for doublet-triplet splitting, iii) a mechanism for generating a $\mathcal{O}(\mathrm{TeV})$ scale $\mu$ parameter, and iv) are consistent with low-energy constraints.

This paper contains two key results. The first is the realization that constraints on bilinear R-parity violation, arising from the upper bound on neutrino masses, provide complementary, stringent, and robust constraints on RPV GUT models compared to those coming from proton-decay (which have traditionally been considered to provide the most stringent constraints on RPV GUT models). The second and more important result is that in currently known UV-motivated $SU(5)$ GUT models satisfying the phenomenological features mentioned above, \emph{constraints on bilinear R-parity violation disfavor any R-parity violation altogether}. This result arises due to the difficulty from a UV point of view in achieving a sufficient hierarchy between the two bilinear superpotential terms $\int d^2\theta \mu\, H_u H_d$ and $\int d^2\theta \kappa_i \, L_i H_u$. For concreteness we frame our arguments within the context of two realistic UV frameworks: i) Heterotic orbifold compactifications and ii) $M$ theory compactifications on $G_2$-manifolds.

In this sense, the arguments in the paper provide a \emph{constrained} top-down argument for R-parity conservation. The constraints arising from bilinear R-parity violation have of course been known for quite some time \cite{Banks:1995by}. However, we believe that the \emph{generality} of the results obtained, especially within the context of UV-motivated phenomenologically viable GUT models, has not been appreciated thus far. For the majority of the paper, we will present our arguments within the context of $SU(5)$ GUT models. Towards the end, we also consider minimal $SO(10)$ GUTs, and argue that the two qualitative results above should remain unchanged. 

The paper is organized as follows. Section \ref{susygut} summarizes ways in which $SU(5)$ GUT models with RPV can avoid fast proton decay. Section \ref{bilinear} reviews the constraints on $\kappa$ from neutrino masses. Section \ref{bilinearSU(5)} gives arguments for why bilinear RPV is always present in realistic $SU(5)$ GUT theories. Section \ref{kappamu} argues that if RPV is allowed in certain viable top-down frameworks, $\kappa/\mu$ will violate neutrino mass bounds. Section \ref{other} discusses RPV in minimal $SO(10)$ models, and we present our conclusions in Section \ref{conclusions}.

\section{Suppressing Proton Decay in RPV SUSY GUTs}\label{susygut}

The renormalizable R-parity violating (RPV) operators in the MSSM superpotential along with the $\mu$ term are given by:\begin{equation}\label{WRPV}
W_{RPV} \supset -\mu H_d H_u - \kappa_i L_i H_u + \lambda_{i j k} L_i L_j E^c_k + \lambda^\prime_{i j k} L_i Q_j D^c_k + \lambda^{\prime \prime}_{i j k} D^c_i D^c_j U^c_k \end{equation} where $i, j, k$ represent generation indices and $\lambda_{i j k}$ and $\lambda^{\prime \prime}_{i j k}$ are antisymmetric under $i \leftrightarrow j$ due to the SM gauge symmetries. In $SU(5)$ symmetric theories with R-parity violation, all three trilinear R-parity violating operators in (\ref{WRPV}) come from the same $SU(5)$ symmetric coupling, $\eta\, {\bf 10_i \overline{5}_j \overline{5}_k}$ where $\eta$ is a dimensionless coupling. The simultaneous presence of $B$ violating and $L$ violating operators induce proton decay, and thus the proton decay bound $\left|\lambda \, \lambda^{\prime \prime}\right| \lesssim 10^{-24} (\tilde{m}/\mathrm{TeV})^2$ \cite{Hinchliffe:1992ad} requires $\eta \lesssim 10^{-12}$ assuming a common squark mass $\tilde{m} \sim \mathcal{O}(\mathrm{TeV})$. 

However, there have been many proposals in the literature which can reconcile R-parity violation in SUSY GUT's with the proton decay bound on $B$ and $L$ violating couplings:\begin{itemize}

\item \emph{Hierarchy in trilinear RPV couplings from $SU(5)$ breaking.} If the trilinear RPV couplings are generated from $SU(5)$ breaking effects, it is possible to establish a hierarchy between $B$ and $L$ violating terms. $SU(5)$ GUT models with predominantly leptonic trilinear RPV couplings have been explored in \cite{Smirnov:1995ey,Tamvakis:1996dk,Barbieri:1997zn,Giudice:1997wb}, while $SU(5)$ GUT models generating predominantly baryonic RPV couplings have been explored in\cite{Smirnov:1995ey,Tamvakis:1996np,DiLuzio:2013ysa}.

\item \emph{Suppressed couplings to light quarks ($u, d, s$).} If RPV couplings involve predominantly heavy quarks ($c, b, t$), the effective operators which induce proton decay will be generated at loop level, suppressed by off-diagonal elements of the CKM matrix \cite{Smirnov:1995ey,Smirnov:1996bg}. This leads to weaker bounds $\left|\lambda^\prime\, \lambda^{\prime\prime}\right| \lesssim \mathcal{O}(10^{-16}) - \mathcal{O}(10^{-10})\, (\tilde{m}/\mathrm{TeV})^2$, depending on the particular generation indices of the couplings.

\item \emph{Heavy $\gtrsim$ PeV scalars.} Diagrams which induce proton decay involve virtual squark exchange; thus the lower bounds on $\left|\lambda^\prime \, \lambda^{\prime \prime}\right|$ scale as $\tilde{m}^2$ where $\tilde{m}$ is the squark mass scale. Thus in scenarios with scalars heavier than a PeV \cite{Wells:2004di,ArkaniHamed:2004fb,Giudice:2004tc}, both the $\lambda^\prime$ and $\lambda^{\prime \prime}$ couplings may not have to be extremely tiny, while still being consistent with proton decay bounds \cite{Gupta:2004dz}.

\item \emph{Tiny trilinear couplings.} Tiny $\lesssim 10^{-15}$ trilinear RPV couplings can be motivated for example by forbidding $\lambda, \lambda^\prime, \lambda^{\prime \prime}$ with a symmetry and generating them from higher-dimensional superpotential or K\"ahler potential operators. An example of the latter case is given in \cite{Kuriyama:2008pv}. In either case this can result in trilinear RPV operators which are suppressed by $\sim\mathrm{TeV}/\Lambda$, where $\Lambda$ could be the Planck scale. The RPV phenomenology will then be dominated by the bilinear $L H_u$ term.

\end{itemize} 
In the cases mentioned above, the proton lifetime is no longer a powerful constraint in ruling out large regions of parameter space. However, we will argue that neutrino mass constraints provide a complementary constraint for these models, due to the unavoidable presence of the bilinear $\kappa  L H_u$ term. As we will discuss in Section \ref{bilinear} the constraints which neutrino masses place on bilinear R-parity violation are more robust than constraints which arise from proton decay. In particular, the neutrino mass constraints on $\kappa$ can not be decoupled by making superpartners arbitrarily heavy, if one wants to be consistent with gauge coupling unification with an MSSM spectrum.

\section{Bilinear R-parity Violation $\&$ Neutrino Mass constraints}\label{bilinear}

It is well known that the bilinear R-parity violating $\int d^2\theta \kappa_i\, L_i H_u$ term can induce non-zero sneutrino {\it vev}'s $\left< \tilde{v}_i\right>$\cite{Aulakh:1982yn,Hall:1983id} which will induce neutrino-neutralino mixing. Upon diagonalizing the resulting  7 $\times$ 7 neutralino-neutrino mass matrix ${\bf M}_\mathcal{N}$, one neutrino species will obtain a tree-level majorana mass, while the other non-zero masses will be generated by loop effects\cite{Mohapatra:1986aw,Banks:1995by,Nowakowski:1995dx,Hempfling:1995wj,deCarlos:1996du,Nardi:1996iy,Chun:1998gp,Chun:1999bq,Hirsch:2000ef}. Combining the recent data from Planck with data on baryon acoustic oscillations, a strict bound has been placed on masses of stable neutrino species: $\sum_{i} m_{\nu_i} \lesssim 0.23$ eV\cite{Ade:2013zuv}. If this bound is taken seriously\footnote{It is possible to evade these bounds if at least one of the SM neutrinos is unstable \cite{Chikashige:1980qk}, or if assumptions regarding $\Lambda_{CDM}$ cosmology are changed. We will not consider these possibilities.}, then neutrino mass bounds place very strict constraints on the bilinear coupling $\kappa$. In this section, we briefly review the resulting constraints on $\kappa$. 

In the presence of bilinear R-parity violation, there are no conserved quantum numbers which can distinguish the $H_d$ superfield from the lepton superfields $L_i$. As a result, there is no unique basis in which $\kappa_i$ and $\mu$ are defined. The bounds on bilinear R-parity violation from neutrino masses is usually stated in terms of a basis independent angle $\xi$, which parameterizes the misalignment between the superpotential parameters\footnote{One can also define a mixing angle $\zeta$ which parameterizes the misalignment between the soft breaking bilinear parameters $V_{soft}\supset B_i L_i H_u$ and the sneutrino/Higgs \emph{vev}s. A non-zero $\zeta$ will induce neutrino masses at loop level \cite{Grossman:1998py,Grossman:1999hc,Davidson:2000uc,Davidson:2000ne}. However, this contribution is more model dependent and can become negligble if one assumes a sufficiently decoupled MSSM Higgs sector \cite{Grossman:2003gq}.} $\mu, \kappa_i$ and the \emph{vev}s $\left<v_d\right>, \left<\tilde{v}_i \right>$\cite{Banks:1995by}: \begin{equation}\cos\xi \equiv \frac{1}{\left|\vec{\mu}\right| \left| \vec{v} \right|} \left(\sum_\alpha \kappa_i  \left<\tilde{v}_i \right> + \mu\,\left<v_d\right>\right). \end{equation} where $\left| \vec{\mu} \right| = \sqrt{\sum_i \kappa_i^2 + \mu^2}$ and $\left| \vec{v} \right| = \sqrt{\sum_i \left<\tilde{v}_i\right>^2 + \left<v_d\right>^2}$. Integrating out the heavier neutralinos induces a rank 1 majorana mass matrix for the neutrinos, resulting in a tree-level contribution to $m_\nu$\cite{Banks:1995by,Nowakowski:1995dx,Hempfling:1995wj}:\begin{equation}\label{neutmass} m_{\nu} \simeq \frac{M_Z^2 \cos^2 \beta (M_1 c_W^2 + M_2 s_W^2)}{M_1 M_2 \left|\vec{\mu}\right| - \sigma M_Z^2 \sin{2\beta} (M_1 c_W^2 + M_2 s_W^2)}\left|\mu\right| \xi^2.\end{equation} In the above, we have taken the small $\xi$ limit which is self-consistent with the neutrino mass bounds. $\sigma$ is the sign of $\mu$ in the basis where $\kappa_i = 0$, and $M_1$ and $M_2$ are the usual soft-breaking gaugino masses. Making the simplifying assumption $\left|M_1\right| = \left|M_2 \right| = \left|\mu\right| = M$ and taking the large $\tan\beta$ limit, we can obtain a limit on $\xi$ from the constraint $m_{\nu} < 0.23$ eV\cite{Ade:2013zuv}:\begin{equation}\label{xiconstraint} \xi \lesssim \frac{10^{-6}}{\cos\beta} \times \left|\frac{M}{M_Z} - \sigma \sin{2\beta} \frac{M_z}{M}\right|^{1/2}.\end{equation} To be consistent with this limit on $\xi$, the four vectors $(\mu, \kappa_i)$ and $(v_d, \tilde{v}_i)$ must be nearly parallel. Without loss of generality, if $\xi \ll 1$ we can always go to a basis where $\vec{\mu} = (\mu, \kappa_i)$ and $\vec{v} = (v_d, \tilde{v}_i)$ such that $\mu \gg \kappa_i$ and $v_d \gg \tilde{v}_i$. In such a basis, the mixing angle $\xi$ can be approximated as:\begin{equation}\label{xiform} \xi = \sqrt{\sum_i \left(\kappa_i/\mu - \tilde{v}_i/v_d\right)^2}.\end{equation} 

If the soft-breaking Lagrangian is of the form\footnote{If off-diagonal soft masses of the form $m_{H_d L_i}^2 H_d^* \tilde{L}_i$ are present, they will also contribute to $\xi$. These off-diagonal terms will be radiatively generated in the presence of trilinear $L$-violating couplings.} \begin{equation}V_{soft} = m_{H_d}^2 \left|H_d\right|^2 + \sum_i m_{L_i}^2 \left|L_i\right|^2 + B_0\, \mu H_u H_d + B_i \, \kappa_i \,H_u L_i + h.c. + ..., \end{equation} then (\ref{xiform}) can be expressed in terms of $\kappa_i$ and the soft parameters as:\begin{equation}\label{xieqn}\xi \simeq \left\{ \sum_i \left(\frac{\kappa_i}{\mu}\right)^2 \left(\frac{\mu \Delta B_i \tan\beta + \Delta m_i^2}{m_{L_i}^2 + M_Z^2 \cos 2\beta/2}\right)^2 \right\}^{1/2}\end{equation} where $\Delta B_i = B_i - B$ and $\Delta m_i^2 = m_{L_i}^2 - m_{H_d}^2$. One can immediately see that if the soft parameters are arbitrary and ${\cal O}(1)$ different from each other,  then $\xi \sim \mathcal{O}(1) \sum_i \left(\left|\kappa_i\right|/\left|\mu\right|\right)^2 $ .

If, on the other hand, supersymmetry breaking is mediated in a \emph{universal} manner with mSUGRA-like boundary conditions, $\Delta B_i$ and $\Delta {m_i}^2$ might vanish at the SUSY breaking messenger scale $\Lambda$. However, the parameters which enter into (\ref{neutmass}) should be renormalized down to the neutralino mass scale $M$\cite{Hempfling:1995wj,Nilles:1996ij}. Assuming $\Delta B_i = 0, \,\Delta {m_i}^2=0$  at the scale $\Lambda = 10^{14} M$, RG effects induce a non-zero $\xi  \sim 10^{-3}/\cos^2\beta \sqrt{\sum_i (\kappa_i^2/\mu^2)}$ upon renormalization to the scale $M$\footnote{$\Delta B_i$ and $\Delta {m_i}^2$ are radiatively driven predominantly by the anomalous dimension of $H_d$, so $\Delta B_i$ and $\Delta {m_i}^2$ are to a good approximation flavor independent. This also explains the presence of $\cos\beta$ in (\ref{xiconstraintfinal}).}. Combining this with (\ref{xiconstraint}) places a bound on $\kappa_i$ in a basis where $\left|\mu\right| \gg \left|\kappa_i\right|$:\begin{equation}\label{xiconstraintfinal} \sqrt{\frac{\sum_i \kappa_i^2}{\mu^2}}  \lesssim 10^{-3}\cos\beta \left|\frac{M}{M_Z} - \sigma \sin{2\beta} \frac{M_z}{M}\right|^{1/2}\end{equation}
Therefore even in the case where soft-parameter universality is motivated by some UV boundary conditions, radiative corrections will induce a non-zero $\xi$, putting significant constraints on the bilinear RPV superpotential terms. 

Not only are the constraints very stringent, they are also rather robust. It is very difficult to decouple the constraint (\ref{xiconstraintfinal}) if one requires consistency with gauge coupling unification. This is because precision gauge coupling unification in minimal $SU(5)$ requires that the Higgsino and Wino masses are near the electroweak scale \cite{ArkaniHamed:2004fb}, unless one adds light, exotic vector-like fermions to the theory \cite{Giudice:2004tc}. Thus if one takes precision gauge coupling unification seriously, $M_1$, $M_2$ and $\mu$ should not be too far from the electroweak scale, making the bound (\ref{xiconstraintfinal}) very difficult to avoid\footnote{Because the Bino does not contribute to running of gauge couplings, taking $M_1 \gg M_Z$ would still maintain gauge coupling unification. However, this is not radiatively stable as $M_1$ contributes to the renormalization of $M_2$ at two-loop \cite{Martin:1993zk}.}, without motivating a hierarchy between the $\kappa_i$ and $\mu$. In contrast, bounds on lepton violation from other $\left|\Delta L\right| = 2$ processes such as $\mu \rightarrow e$ conversion\cite{Lee:1984kr,Lee:1984tn} and $K^+ \rightarrow \pi^- \ell^+ \ell^-$\cite{Littenberg:1991ek,Littenberg:2000fg} decouple in the limit of heavy scalar superpartners, and are therefore less robust. In the next section we will argue that the bilinear $\int d^2\theta \kappa_i\, L_i H_u$ term is present in all known $SU(5)$ GUT models with R-parity violation which satisfy t'Hooft's criterion of naturalness.

\section{Ubiquity of Bilinear R-parity Violation in $SU(5)$ GUTs}\label{bilinearSU(5)}

In this section we argue that in realistic minimal $SU(5)$ GUT theories with R-parity violation, the bilinear $\int d^2\theta \kappa_i\, L_i H_u$ is always present in the effective Lagrangian. A key point in this argument is that for realistic MSSM theories, \emph{there are no symmetries which can protect $\kappa = 0$ in the presence of trilinear leptonic RPV}. The argument goes as follows. Suppose that there is an exact global symmetry $\mathcal{H}$ in the MSSM Lagrangian. From the arguments of \cite{Leurer:1992wg}, $\mathcal{H}$ must be flavor blind, or else there will be either degenerate fermion masses and/or zeros in $V_{CKM}$ and $U_{PMNS}$. Assuming no exotic matter content, this constrains $\mathcal{H}$ to be flavor-blind and Abelian. Requiring that the lepton and down-type Yukawa couplings are $\mathcal{H}$ invariant leads to the following relations amongst $\mathcal{H}$ charges:\begin{equation}\label{chargerelations} Q_{L} + Q_{H_u} = Q_{Q} + Q_{L} + Q_{D^c} + Q_{\mu} = Q_{L} + Q_L + Q_{E^c} + Q_{\mu},
\end{equation} where $Q_\mu \equiv Q_{H_u} + Q_{H_d} - 2 q_\theta$ and we have allowed for cases in which $\mathcal{H}$ is an R-symmetry: $q_\theta \neq 0$. Thus if any of the trilinear leptonic RPV couplings along with the $\int d^2\theta \mu\, H_u H_d$ term are $\mathcal{H}$ invariant, $\int d^2\theta \kappa L H_u$ will also be $\mathcal{H}$ invariant. 

Without a symmetry protecting $\kappa = 0$, there will be K\"{a}hler potential contributions to $\kappa$ which are not protected by non-renormalization. For instance, wave-function renormalization diagrams will induce mixing between $L$ and $H_d$, resulting in contributions of order $\kappa \sim \lambda \,y_\tau \mu /16 \pi^2$, $ \kappa \sim \lambda^\prime y_b \mu/16 \pi^2$\cite{deCarlos:1996du,Nardi:1996iy}. A more dangerous source of K\"{a}hler potential corrections arises once SUSY breaking is mediated to the visible sector via terms of the form:\begin{equation}\label{kahlercorrections}
K \supset \alpha \frac{X^\dagger}{M} L H_u + \beta \frac{X^\dagger}{M} H_u H_d + h.c.
\end{equation} where $X$ is a SUSY breaking spurion and $M$ is the messenger scale. This leads to contributions of order $\kappa \sim \mu \sim \left<F_X\right>/M$. The only way to avoid such contributions in theories with RPV is to suppose that only the baryonic R-parity violating $\int d^2\theta \, U^c D^c D^c$ term is allowed in the Lagrangian. However, this requires that $\mathcal{H}$ does not commute with $SU(5)$. This leads to the following considerations.

In realistic $SU(5)$ GUTs, some GUT breaking mechanism must be present to break $SU(5)$ to $G_{SM}$ and solve doublet-triplet splitting. One might imagine that including these effects can allow for a symmetry $\mathcal{H}$ which does not commute with $SU(5)$. In the following, we examine this question in the context of two UV frameworks which contain natural mechanisms for GUT breaking and doublet-triplet splitting: i) Heterotic orbifold compactifications\footnote{As we will discuss, our arguments here also apply to more general field-theoretic orbifold GUTs .} and ii) $M$ theory compactifications on $G_2$-manifolds. We also discuss ``bottom up" models mentioned in Section \ref{susygut} which generate R-parity violation via GUT breaking effects. In all cases, we argue that no exact symmetries can forbid contributions like (\ref{kahlercorrections}) in the presence of other RPV operators.

\subsection{Heterotic Orbifold Compactifications}\label{heterotic}

One top-down GUT framework which can achieve realistic phenomenology arises from orbifold compactifications of the Heterotic string \cite{Dixon:1985jw,Dixon:1986jc,Ibanez:1986tp,Ibanez:1987sn,Faraggi:2001ry}. These frameworks rely on a mechanism first proposed by Witten in which nontrivial Wilson lines simultaneously break the GUT symmetry while projecting out Higgs triplet zero modes to achieve doublet-triplet splitting\cite{Witten:1985xc}. Analyses of the Heterotic $E_8 \times E_8$ string compactified on toroidal orbifolds have demonstrated numerous vacua with realistic MSSM spectra and R-parity conservation\cite{Lebedev:2006kn,Lebedev:2007hv,Lebedev:2008un}. However, the possibility of phenomenologically viable models with $R$-parity violation in this framework has not yet been fully explored. Here we simply consider the features of Heterotic orbifold models already present in the literature, and discuss implications with respect to R-parity violation.

Global symmetries $\mathcal{H}$ in this framework can arise either from a subgroup of $E_8 \times E_8$ and/or geometric orbifold selection rules. In the presence of non-trivial orbifold boundary conditions, the same orbifold projection which solves triplet-doublet splitting in the Higgs sector will generically give rise to massless split GUT multiplets in the matter sector as well; the $\mathcal{H}$ quantum numbers in these sectors need not commute with $SU(5)$. However, certain twisted sector states do not feel the effects of GUT-breaking Wilson lines, and massless states corresponding to these ``local GUT" sectors give rise to complete GUT multiplets. As a result, the $\mathcal{H}$ quantum numbers of matter arising from these sectors must commute with $G_{GUT}$. Suppose a single generation of SM matter arises from a complete GUT multiplet, while other generations arise from split GUT multiplets. If the $\mathcal{H}$ quantum numbers of only one generation commutes with $G_{GUT}$, then $\mathcal{H}$ would by definition be a flavor symmetry, which is inconsistent with measurements of $V_{CKM}$ and $U_{PMNS}$ \cite{Leurer:1992wg}. 

\emph{Thus if at least one generation of SM matter comes from a complete GUT multiplet, $\mathcal{H}$ must commute with $G_{GUT}$ in the matter sector if $\mathcal{H}$ is unbroken, even if other generations come from split GUT multiplets.} This is certainly the case for the Heterotic orbifold models referenced above, as well as the class of orbifolds studied more recently in \cite{Blaszczyk:2009in}. These arguments also apply more generally to purely field-theoretic orbifold GUTs\cite{Kawamura:2000ev,Hall:2001pg,Hebecker:2001wq} which solve doublet-triplet splitting with a similar mechanism.

\subsection{$M$ Theory Compactifications on Singular $G_2$-Manifolds}\label{M-theory}

Another, related, realistic top-down GUT framework is given by $M$ theory compactified on singular seven-dimensional manifolds with $G_2$ holonomy\cite{th/9506150,th/0011089,th/0107177,th/0109152,th/0701034,0801.0478}. Like in the Heterotic orbifold models, breaking $G_{GUT}$ to $G_{SM}$ in $M$ theory compactifications can be mediated by a Wilson line $W$ which commutes with $G_{SM}$. As was first noted by Witten\cite{ph/0201018}, the presence of non-trivial $W$ allows the geometric construction of discrete symmetries which do not commute with $G_{GUT}$, even if the corresponding vacuum with trivial $W$ is fully GUT symmetric. This discrete symmetry can forbid the Higgs doublet mass term while allowing the corresponding triplet mass term, solving doublet-triplet splitting. In what follows we briefly review the relevant details of this construction; interested readers should consult \cite{ph/0201018} for additional details.

In realistic $M$ theory compactifications, non-Abelian gauge fields are localized on a 3-dimensional submanifold $Q$ within the $G_2$ manifold, while chiral fermions are localized at isolated points within $Q$\cite{th/0109152}. $Q$ may admit the action of some discrete symmetry $\mathcal{H}^\prime$; following the example in \cite{ph/0201018} we take $\mathcal{H}^\prime \cong F \cong Z_N$ where the group of incontractible loops on $Q$ is $\pi_1\left(Q\right) = F$. In the presence of a Wilson line background $W$, $\mathcal{H}^\prime$ will also act non-trivially on the gauge bundle over $Q$. Suppose there are two $\mathcal{H}^\prime$ orbits in $Q$, $S_1$ and $S_2$, such that $\mathcal{H}^\prime$ acts trivially on the gauge fibers of $S_1$ and acts with a gauge transformation by $W$ on the gauge fibers of $S_2$. The commutant of $G_{SM}$ within $SU(5)$ is $U(1)_Y$; this fixes $W = Diag(e^{4 \pi i \rho/N},e^{4 \pi i \rho/N},e^{4 \pi i \rho/N}, e^{- 6 \pi i \rho/N}, e^{- 6 \pi i \rho/N})$ with integer $\rho$. Consequently, the $\mathcal{H}^\prime$ charges of chiral matter satisfy:
\begin{equation}\label{Wilsonline}
Q_{i} = Q_{SU(5)} + \delta_i\, Q_Y
\end{equation}
where $Q_Y$ is the hypercharge of a given field, normalized such that the quark doublet has $Q_Y = 1$. If the $i$'th superfield is localized on $S_1$ then $\delta_i = 0$, otherwise if it is localized on $S_2$ then $\delta_i = 1$. By localizing ${\bf 5}_{H_u}$ on $S_1$ and ${\bf \overline{5}}_{H_d}$ on $S_2$ or vice versa, doublet-triplet splitting is achieved if $Q_{SU(5)}$ for ${\bf 5}_{H_u}$ and $\bf \overline{5}_{H_d}$ are chosen such that the triplet mass term is allowed. This in turn fixes the charge of the $H_u H_d$ term to be:\begin{equation}\label{tripdoublet}
Q_{H_u} + Q_{H_d} = 5\, \rho\, (\delta_{5_{H_u}} - \delta_{ \overline{5}_{H_d}}) + 2 q_\theta
\end{equation} where we have introduced a $\theta$ charge $q_\theta$ to include $R$-symmetries. Thus if $5 \rho \neq 0\, mod\, N$, the triplet mass term is allowed while the doublet term is forbidden \cite{ph/0201018}. 

However, realistic models require that this symmetry is broken to generate a non-zero $\mu$ term. Therefore this $\mathcal{H}^\prime \cong Z_N$ symmetry must be broken by a \emph{vev} with charge $\mp 5 \rho\, mod\, N$. As a result, $\mathcal{H}^\prime$ will be broken to a subgroup $\mathcal{H}\cong Z_M$ with $ 5 \rho = 0\, mod \, M$. It is straightforward to see from (\ref{Wilsonline}) that the action of $W$ splits the $Z_N$ charges within $SU(5)$ multiplets by $\pm 5 \rho$. Thus once a non-zero $\mu$ term is generated, $\mathcal{H}^\prime$ will be broken to a symmetry $\mathcal{H}$ which commutes with $SU(5)$ in the MSSM Lagrangian. Note that a priori, $\mathcal{H}$ may or may not be trivial.  

\subsection{K\"{a}hler Potential Corrections in ``Bottom-Up" GUT Models}\label{SU(5)general}

Here we briefly discuss bilinear R-parity violation in $SU(5)$ GUT models where RPV operators are generated by $SU(5)$ GUT breaking effects, many of which were referred to in Section \ref{susygut}. The GUT models in \cite{Smirnov:1995ey,Tamvakis:1996dk,Barbieri:1997zn,Giudice:1997wb} result in theories with leptonic R-parity violation while baryonic R-parity violation is either absent or suppressed; the earlier arguments of this section then imply that the $\kappa$ terms will recieve corrections from K\"{a}hler potential operators. There have also been $SU(5)$ GUT models proposed that generate a superpotential with predominantly baryonic R-parity violation via the missing partner mechanism \cite{Smirnov:1995ey,Tamvakis:1996np,DiLuzio:2013ysa}. The superpotentials for these models are \emph{non-generic}, in that they do not include all terms which are consistent with the global symmetries of the theory. In fact, for the $SU(5)$ GUT models in \cite{Smirnov:1995ey,Tamvakis:1996np}, as well as several of the $SO(10)$ GUT models proposed in \cite{DiLuzio:2013ysa}, there is no global symmetry that distinguishes the matter ${\bf \overline{5}}_M$ from the Higgs ${\bf \overline{5}}_{H_d}$. Thus in the presence of a non-vanishing $\mu$ term, there will generically be K\"{a}hler potential operators such as (\ref{kahlercorrections}) giving rise to an effective $\kappa$ term. With the exception of \cite{Giudice:1997wb}, the importance of these K\"{a}hler potential corrections has largely been ignored in the literature mentioned above\footnote{The importance of K\"{a}hler potential corrections to RPV operators was also recently emphasized in \cite{Csaki:2013jza}.}. Whether or not such contributions violate the bounds in Section \ref{bilinear} is a model dependent question, which must be addressed for any realistic GUT theory with R-parity violation. From a string/$M$ theoretic point of view it seems difficult to implement these models without generating large K\"{a}hler potential corrections to $\kappa$.

\section{$\kappa/\mu$ in Top-Down $SU(5)$ GUTs} \label{kappamu}

In the previous section, we argued that the bilinear $\int d^2\theta \kappa\,L H_u$ will always be present in realistic $SU(5)$ GUTs. However, (\ref{xiconstraintfinal}) imposes meaningful constraints only for frameworks in which a mechanism exists for generating phenomenologically viable $\mu$ (and the associated $B\mu$) parameters. In general, these mechanisms depend on the nature of supersymmetry breaking and its mediation. Gauge mediation models generically lead to a ``$\mu/B\mu$ problem"\cite{Dvali:1996cu}, while pure anomaly mediation leads to tachyonic sleptons\cite{Randall:1998uk}. On the other hand, gravity mediation is free from these issues, and can naturally combined with a GUT framework arising in string theory compactifications, such as the Heterotic orbifold and $M$ theory compactifications discussed in Section \ref{bilinearSU(5)}. Therefore in this section we will study the magnitude of $\kappa/\mu$ in the top-down GUT frameworks discussed in Section \ref{bilinearSU(5)}, with gravity mediated supersymmetry breaking. 

From an effective supergravity point of view, there are two elegant mechanisms for achieving a phenomenologically viable $\mu$ term:

\begin{itemize}
\item {\bf The Kim-Nilles/Casas-Munoz (KN/CM) mechanism}\cite{Kim:1983dt,Casas:1992mk}. In this case, the $\mu$ term is generated by a higher dimensional superpotential operator which can be naturally small. An interesting example is $\int d^2 \theta\,\frac{\left<W_0\right>}{M_{pl}^2}\,H_u H_d$ where $\left<W_0\right>$ is the \emph{vev} of a hidden sector superpotential. If $\left<W_0\right>$ is the dominant source of SUSY breaking, then $m_{3/2} \sim \left<W_0\right>/{M_{pl}}^2$ which results in $\mu \sim m_{3/2}$.
\item {\bf The Giudice-Masiero (GM) mechanism}\cite{Giudice:1988yz}. In this class of mechanisms, the $\mu$ term is generated by K\"{a}hler potential operators of the form $\int d^4 \theta\,\alpha \left(\frac{X^\dagger}{2 M_{pl}}\right)\, H_u H_d+ h.c.$. Decoupling gravity by taking the ``flat limit" $M_{pl} \rightarrow \infty$ with $m_{3/2}$ fixed leads to a global SUSY theory with an effective superpotential term $\mu = \alpha \, m_{3/2}$.
\end{itemize} 

Both mechanisms involve a symmetry $\mathcal{H}^\prime$ which forbids $\int d^2\theta \mu\, H_u H_d$ that is spontaneously broken in order to generate non-zero $\mu$. We will see below that Heterotic orbifold compactifications can naturally realize the KN/CM mechanism while the M-theory compactifications can naturally realize the GM mechanism. However, unless a symmetry is present which forbids all RPV operators, these mechanisms will also generate $\kappa$ in accordance with the arguments of Section \ref{bilinearSU(5)}. We will show below that in these frameworks, if RPV is allowed then $\kappa$ will be generated with $\kappa/\mu \gg 10^{-3}$. Therefore, we find that R-parity violation in these frameworks is disfavored by neutrino mass bounds, leaving R-parity conservation as the only viable alternative.

In the following, our focus is on the case where the symmetry which forbids $\int d^2\theta \kappa L H_u$ is flavor blind. If instead a flavor symmetry forbids $\int d^2\theta \kappa\, L H_u$, it must be spontaneously broken to generate off-diagonal elements in both $V_{CKM}$ and $U_{PMNS}$\cite{Leurer:1992wg}. However, because $U_{PMNS} \gg V_{CKM}$ in the off-diagonal elements, it seems rather difficult to achieve realistic $V_{CKM}$ and $U_{PMNS}$ via a spontaneously broken symmetry which commutes with $SU(5)$, which at the same time also generates a viable $\mu$ and sufficiently suppressed $\kappa$. We will not consider this possibility henceforth. 

\subsection{$\kappa/\mu$ in Heterotic Orbifold Compactifications}\label{heterotic}

As argued in Section \ref{heterotic}, a flavor-blind global symmetry $\mathcal{H}^\prime$ will commute with $SU(5)$ in realistic Heterotic orbifold compactifications. If $\mathcal{H}^\prime$ has a role in solving the $\mu$ problem, $SU(5)$ anomaly universality requires \cite{1009.0905,1102.3595} that $\mathcal{H}$ is a discrete R-symmetry, $\mathcal{H}^\prime \cong Z_M^R$. An elegant solution to the $\mu$ problem in Heterotic orbifold compactifications can then occur\cite{Lebedev:2006tr,Kappl:2008ie,Brummer:2010fr} if $Z_M^R$ is broken non-perturbatively by hidden sector gaugino condensation\cite{Affleck:1983mk}, which generates a non-perturbative superpotential $\left<W_0\right> \sim M_{pl}^3\,e^{-b\,S}$ for the complex dilaton $S$. The KN/CM mechanism can then be naturally implemented if $H_u H_d$ is uncharged under $Z_M^R$ such that $\int d^2\theta\, W_0$ and $\int d^2 \theta \,W_0 \,H_u  H_d/ M_{pl}^2$ are $Z_M^R$ invariant\footnote{This also allows Guidice-Masiero contributions to $\mu$. One could imagine situations, however, where the dominant contribution arises from the KN/CS mechanism.}. In gravity mediation, this results in $\mu \sim \left<W_0\right>/M_{pl}^2 \sim m_{3/2}$ if $\left<W_0\right>$ is the dominant source of supersymmetry breaking (as in dilaton dominated supersymmetry breaking). 

It was shown in \cite{1102.3595,Chen:2012jg} that for $M \le 36$, only two $Z_M^R$ symmetries which satisfy $SU(5)$ anomaly universality and allow $\int d^2\theta\, W_0 \, H_u H_d$ also allow for R-parity violation once $Z_M^R$ is broken by $\left<W_0\right>$. They are listed in Table \ref{table:z4r}. $Z_4^R(I)$ allows the bare $\int d^2\theta \kappa L H_u$ term, which is a disaster since there is nothing protecting $\kappa \sim M_{GUT}$. On the other hand, $Z_4^R(II)$ has the feature that both $L\,H_u$ and $H_u\,H_d$ have the same charge. Thus,  $\langle W_0\rangle$ generates both the $\mu$ and $\kappa$ terms by the KN/CM mechanism, resulting in $\kappa \simeq \mu$. This then leaves $Z_M^R$ symmetries which conserve R-parity in the presence of non-vanishing $\left<W_0\right>$ as the only phenomenologically viable option (at least for $M \le 36$). We expect similar results for higher order symmetries as well.

\begin{table}[t!]
\centering
\begin{tabular}{|c| c | c | c | c | c |}
\hline
& $Q_{\bf 10}$ & $Q_{\bf \overline{5}}$ & $ Q_{H_u}$  & $ Q_{H_d}$ & $q_\theta$ \\
\hline
$Z_4^R( I)$ & 0 & 0 & 2 & 2 & 1\\
\hline
$Z_4^R(II)$ & 2 & 2 & 2 & 2 & 1\\
\hline
\end{tabular}
\caption{$Z_4^R$ symmetries which satisfy $SU(5)$ anomaly universality and solve the $\mu$ problem while allowing MSSM Yukawa couplings and R-parity violation once a non-zero $\mu$ term is generated. These symmetries were first given in Table 1 of \cite{Chen:2012jg}.}
\label{table:z4r}
\end{table}

In principle, additional symmetries which forbid $\kappa$ may arise from subgroups of $E_8 \times E_8$ or orbifold selection rules. Such symmetries are typically broken by $D$ and $F$-flatness conditions\cite{Cleaver:1997jb} required to cancel stringy Fayet-Illiopoulos D-terms associated with anomalous $U(1)$'s\cite{Dine:1987xk,Atick:1987gy}. This involves giving \emph{vev}s to numerous SM singlets with $10^{-2} \lesssim \left<\phi\right>/M_{pl} \lesssim 10^{-1}$, so suppression by a single factor of $\left<\phi\right>/M_{pl}$ results in $\left|\kappa\right|/\left|\mu\right| \gtrsim 10^{-2}$ which is not sufficient to avoid the bounds of Section \ref{bilinear}. Furthermore, these spontaneously broken symmetries must allow the $\mu$ term so that $\mu$ is not also suppresed by factors of $\left<\phi_i\right>/M_{pl}$. An example of such a symmetry in Heterotic Calabi-Yau compactifications which is broken by $D$-flatness is given in \cite{Kuriyama:2008pv}. However for the ``4+1" model in \cite{Kuriyama:2008pv}\footnote{The other ``3+2" model in \cite{Kuriyama:2008pv} does not solve the $\mu$ problem.}, there is no symmetry which forbids higher-dimensional K\"{a}hler potential operators of the form $\int d^4\theta \alpha\left<\overline{N}^c\right>^\dagger L\,H_u/M_{pl}$. In gravity mediation, this will result in $\kappa \sim  \left<\overline{N}^c\right> m_{3/2} /M_{pl}\gtrsim 10^{-2} m_{3/2}$. If one generates a viable $\mu$ parameter with $\mu \lesssim m_{3/2}$, then $\kappa/\mu$ will be too large. Furthermore there is also the issue of the $\mu/B\mu$ problem in \cite{Kuriyama:2008pv}, since both $\mu$ and $B\mu$ are generated at one loop by integrating out GUT-scale particles. 

Thus we argue that it is reasonable to expect that in this class of UV-motivated $SU(5)$ GUT models, bilinear RPV operators, if generated, are generated at a level which is too large. R-parity conservation is then the only viable possibility.

\subsection{$\kappa/\mu$ in $M$ theory Compactifications on $G_2$ Manifolds}

In $M$-theory compactifications on manifolds of $G_2$ holonomy, doublet-triplet splitting is achieved via the symmetry $\mathcal{H}^\prime$ discussed in Section \ref{M-theory}. However,  $\mathcal{H}^\prime$ must be broken in order to generate a non-zero $\mu$ term. It was argued in \cite{1102.0556} that moduli stabilization can break $\mathcal{H}^\prime$ and generate non-zero $\mu$ via the GM mechanism, resulting in $\mu \sim 0.1\, m_{3/2}$. In this section, we argue that if this same mechanism also generates RPV operators, there will also be GM contributions to $\kappa$ which are exlcuded by the neutrino mass constraints of Section \ref{bilinear}. 

Given a $\mathcal{H}^\prime \cong Z_N$ symmetry, the K\"{a}hler potential terms which can contribute to $\mu$ and $\kappa$ upon moduli stabilization are of the form\cite{1102.0556}:\begin{equation}\label{modulikahler}
K \supset \int d^4\theta \,\alpha_1 \left(\frac{\Phi_{q_\mu}}{M_{pl}}\right) H_u H_d + \alpha_2 \left(\frac{\Phi_{q_\kappa}}{M_{pl}}\right) H_u L + h.c., \hspace{5mm} \Phi_{q} \equiv \frac{1}{\sqrt{N}}\left( \sum_{k = 1, N} e^{- 2 \pi i k q/N} s_k \right)
\end{equation} where $q_\mu$ and $q_\kappa$ are such that (\ref{modulikahler}) is $Z_N$ invariant and generically $\alpha_1 \sim \alpha_2$. Here the $s_i$ represent real moduli fields which parameterize areas of 3-cycles in the $G_2$ manifold\cite{th/0203061}. Because they are real, the $s_i$ must transform in a cyclic representation of $Z_N$. One can verify that the $\Phi_q$ form linear representations of $Z_N$, as under the $Z_N$ action $s_1 \rightarrow s_2 \rightarrow ... s_N \rightarrow s_1$, $\Phi_q \rightarrow e^{2 \pi i q/N} \Phi_q$. A $\mu$ term of the correct size is generated when $\left<\Phi_{q_\mu}\right>/M_{pl} \sim 0.1$; satisfying neutrino mass bounds then requires $\left<\Phi_{q_\kappa}\right> \lesssim 10^{-3} \left<\Phi_{q_\mu}\right>$. Therefore to estimate the size of $\kappa$/$\mu$, we must understand the relation between the \emph{vev}s of $\Phi_{q_\kappa}$ and $\Phi_{q_\mu}$.

The key point here is that in the compactified $M$-theory context, for a generic moduli K\"{a}hler potential the real moduli $s_i$ will all obtain \emph{vev}s of the same order\cite{th/0701034}. Upon moduli stabilization, this leads to two possibilities for the $\left<\Phi_q\right>$'s:\begin{enumerate} \item $Z_N$ is completely broken such that $\left<\Phi_q\right> \neq 0$ for all $q$\footnote{For special choices of the moduli K\"{a}hler potential, it may be possible that moduli stabilization results in $\left<\Phi_{q}\right>= 0$ at tree level without the presence of a residual symmetry. However because $\left<\Phi_{q} \right> = 0$ is not protected by any symmetry, it may lifted by loop corrections such as those discussed in \cite{Choi:1997de}.}. Barring a tuning in the \emph{vev}s of $s_i$, this results in all $\left<\Phi_q \right>$'s obtaining \emph{vev}s of the same order such that $\left<\Phi_{q_\kappa}\right> \sim \left<\Phi_{q_\mu}\right>$.
\item Moduli stabilization will leave an unbroken $Z_M$ subgroup of $Z_N$ such that $\left<\Phi_{n M}\right> = 0$  for integer $n$. This occurs if $Z_M$ enforces a relation amongst the $s_i$; for instance a $Z_4$ symmetry acting as $s_1 \rightarrow s_2 \rightarrow s_3 \rightarrow s_4 \rightarrow s_1$ is broken to $Z_2$ if $s_1 = s_3$ and $s_2 = s_4$ but $s_1 \neq s_2$. Generating $\mu$ then requires $q_\mu = 0 \,\, Mod\,\,M$.\end{enumerate}

In both cases if $\left<\Phi_{q_\kappa} \right> \neq 0$, we expect $\left<\Phi_{q_\kappa} \right> \sim  \left<\Phi_{q_\kappa} \right>$ and thus $\mu \sim \kappa$ from (\ref{modulikahler}) which is excluded by neutrino mass constraints. The only remaining option is for a residual $Z_M$ symmetry to enforce $\left<\Phi_{q_\kappa}\right> = 0$; because $Z_M$ commutes with $SU(5)$ this would result in a symmetry which forbids \emph{all} RPV operators (see the discussion in Section \ref{M-theory}). Thus without referring to the specifics of the $Z_N$ symmetry (charge assignments, non-R versus R-symmetry, etc.), we have provided arguments indicating that the possibility of R-parity violation in generic phenomenolgically viable $G_2$ compactifications is disfavored by neutrino mass constraints.

\section{Minimal $SO(10)$ GUT Models}\label{other}

In the previous sections, we provided arguments for the lack of any R-parity violation in well-motivated top-down $SU(5)$ GUT models. Here, we discuss briefly the situation for $SO(10)$ GUT models based on conventional $SU(5) \subset SO(10)$, and argue that given minimal matter content, it is challenging from a GUT point of view to have a viable theory with R-parity violation while respecting neutrino mass constraints. These results only apply to the standard $SU(5)\times U(1)_\chi \supset SO(10)$ embedding. The issue of RPV in flipped $SU(5)$ models is qualitatively different and discussed in \cite{Kuflik:2010dg}.

In the standard $SO(10)$ case, if representations larger than or equal to the {\bf 144} are absent\cite{Martin:1992mq} the only couplings which can give rise to RPV are:\begin{align}\label{SO(10)couplings} \notag W   & \supset \frac{y^{a}_{i j k l}}{\Lambda}{\bf 16}_i \times {\bf 16}_j \times {\bf 16}_k \times {\bf 16}_l + y^{b}_{i j} \, {\bf 16}_i \times {\bf 16}_j \times {\bf 10} \\ & = \frac{y^a_{i j k l}}{\Lambda}\nu^c_i \big(U^c_j D^c_k D^c_l + Q_j L_k D^c_l + E^c_j L_k L_l\big) + y^b_{i j}\, \nu^c_i L_j H_u + ...
\end{align} where $\nu^c_i$ is the SM singlet within the ${\bf 16}_i$ of $SO(10)$, and in the second line we have ommited non-RPV terms. Assuming minimal matter content, the indices run from $i = 1 , 2 , 3$. Thus in minimal $SO(10)$ models, the right-handed sneutrino $\tilde{\nu}^c$ must obtain a \emph{vev} to generate RPV operators.

Now we must consider how a nonzero $\left<\tilde{\nu}^c \right>$ is dynamically generated. Since the $SO(10)$ gauge symmetry forbids any tadpole terms for $\nu^c$, the only remaining possibility is to radiatively induce a tachyonic soft mass for $\tilde{\nu}^c$. Avoiding collider constraints on the $U(1)_\chi$ gauge boson reqires $\left<\nu^c\right> \gtrsim 3$ TeV (assuming $SO(10)$ gauge coupling unification), which implies $y^b_{i j} \lesssim 10^{-3}$ in order to satisfy the weaker bounds given in (\ref{xiconstraintfinal}). Then $\nu^c$ has no sizable couplings in the superpotential, and the only way\footnote{We assume here that the soft breaking trilinears $\mathcal{L} \sim A_{\nu^c}\tilde{\nu}^c \tilde{L} H_u$ can be approximated as $A_{\nu^c}\sim y_{\nu} \tilde{m}$ where $\tilde{m}$ is some common soft mass scale.} to radiatively drive $\tilde{m}_{\nu^c}^2 < 0$ is with a large $S_\chi$ where:\begin{equation}\label{supertrace} S_\chi \equiv Tr( Q_\chi \tilde{m}^2) = 4\left(m_{H_u}^2 - m_{H_d}^2\right)+ Tr\left(6 m_{\tilde{L}}^2 + 9 m_{\tilde{D^c}}^2 -6 m_{\tilde{Q}}^2 - m_{\tilde{E^c}}^2 - 3 m_{\tilde{U^c}}^2- 5 m_{\tilde{\nu}^c}\right)\end{equation} up to the normalization of $Q_\chi$. In the second equality, the trace is taken to be over flavor indices. From (\ref{supertrace}) it is evident that if the soft masses satisfy $SO(10)$ relations $m_{H_d} = m_{H_u} = \tilde{m}_{{\bf 10}}$ and $m_{\tilde{Q}} = m_{\tilde{U}^c} = m_{\tilde{D}^c} = m_{\tilde{L}} = m_{\tilde{E}^c} = m_{\tilde{\nu}^c} = \tilde{m}_{\bf 16}$, $S_\chi = 0$ and is radiatively generated at the two-loop level\cite{Martin:1993zk}. Thus if the soft breaking masses respect $SO(10)$ symmetry at the scale of SUSY breaking, there is no way to radiatively induce a non-zero $\left<\tilde{\nu}^c\right>$ to generate RPV operators unless there is a substantial $\sim 10^{-2}$ hierarchy in the soft scalar masses\footnote{Spontaneous RPV through a non-zero $\left<\tilde{\nu}^c\right>$ has been discussed in $U(1)_{B-L}$ extensions of the MSSM\cite{FileviezPerez:2008sx,Barger:2008wn,Ambroso:2009jd,Ambroso:2010pe,Marshall:2014kea}, motivated by certain Calabi-Yau compcatifications of the Heterotic string\cite{Braun:2005nv}. A large $S_{B-L}$ is required to break $U(1)_{B-L}$, which again requires significant non-universality in the soft scalar masses; the potential origin for such non-universality has not yet been discussed.}.

If somehow a large $S_\chi$ can be generated from $SO(10)$ breaking effects, there is an additional significant problem for $SO(10)$ GUTs in which \emph{all} Yukawa couplings arise from the coupling:\begin{equation} y^b_{i j } {\bf 16}_i \times {\bf 16}_j \times {\bf 10} \rightarrow y^b_{i j} (Q_i H_u U^c_j + Q_i H_d D^c_j + L_i H_d E^c_j + \nu^c_i H_u L_j). \end{equation} In order to satisfy neutrino mass constraints in the presence of a TeV scale $\left< \nu^c \right>$, one must have either $y^b_{i j } \lesssim 10^{-3}$ for all $i, j$ or flavor-dependent soft masses such that only $\left<\tilde{\nu}^c_e\right>$ is nonzero. In the former case, all third generation fermion masses must arise from $SO(10)$ breaking effects. 

These issues represent significant challenges for constructing minimal $SO(10)$ GUT models with phenomenolgically viable RPV. We remark however that these conclusions might be avoided if one adds exotics, for example a ${\bf 16}'$, ${\bf \overline{16}}\,'$ pair whose \emph{vev}s generate RPV couplings.

\section{Conclusions}\label{conclusions}

In this work, we have explored the possibility of phenomenologically viable R-parity violation in $SU(5)$ GUT models motivated from a UV point of view. This restricts us to consider models which have a mechanism of GUT breaking to the SM gauge group, solutions to the doublet-triplet splitting and $\mu$-$B\mu$ problems, and 
broad consistency with low energy constraints such as those from fermion masses and mixings, proton decay, etc. We have shown from our analysis that imposing the above requirements on well-motivated top-down $SU(5)$ GUT models, gives rise to  one of the two situations - a) all R-parity violating operators are present
, or  b) No R-parity violating operators are present. Furthermore, in well-motivated models, it can be shown that in case a) the ratio $\kappa/\mu$ is $\mathcal{O}(1)$ without extreme fine-tuning. The extremely stringent upper bound on this ratio, therefore, precludes case a) as a viable possibility, leaving R-parity conservation as the only allowed possibility. The arguments can be extended for minimal $SO(10)$ GUTs, giving rise to the same qualitative result (although for slightly different reasons).

From a low-energy point of view, of course, it is still possible that R-parity violation is observed at the LHC.  Our results then show that this would disfavor the entire class of top-down GUT models studied in this work.

\section*{Acknowledgments}

PK would like to thank Brent Nelson for helpful discussions. BZ would like to acknowledge Sebastian Ellis and Garrett L. Edell for helpful conversations. The work of GK, RL and BZ is supported by DoE grant DE-FG-02-95ER40899 and by the MCTP. RL is also supported by DoE grant DE-FG-02-95ER40896. The work of PK is supported by DoE grant DE-FG-02-92ER40704. BSA gratefully acknowledges support of the Science and Technology Facilities Council.

\bibliographystyle{utcaps}
\bibliography{bib}

\providecommand{\href}[2]{#2}\begingroup\raggedright\begin{thebibliography}{10}

\bibitem{Salam:1974xa}
A.~Salam and J.~Strathdee
\href{http://dx.doi.org/10.1016/0550-3213(75)90253-9}{{\em Nucl.Phys.}
  {\bfseries B87} (1975) 85}.

\bibitem{Farrar:1978xj}
G.~R. Farrar and P.~Fayet
\href{http://dx.doi.org/10.1016/0370-2693(78)90858-4}{{\em Phys.Lett.}
  {\bfseries B76} (1978) 575--579}.

\bibitem{Hinchliffe:1992ad}
I.~Hinchliffe and T.~Kaeding
\href{http://dx.doi.org/10.1103/PhysRevD.47.279}{{\em Phys.Rev.} {\bfseries
  D47} (1993) 279--284}.

\bibitem{Goldberg:1983nd}
H.~Goldberg
\href{http://dx.doi.org/10.1103/PhysRevLett.50.1419}{{\em Phys.Rev.Lett.}
  {\bfseries 50} (1983) 1419}.

\bibitem{Ellis:1983ew}
J.~R. Ellis, J.~Hagelin, D.~V. Nanopoulos, K.~A. Olive, and M.~Srednicki
\href{http://dx.doi.org/10.1016/0550-3213(84)90461-9}{{\em Nucl.Phys.}
  {\bfseries B238} (1984) 453--476}.

\bibitem{Aulakh:1982yn}
C.~Aulakh and R.~N. Mohapatra
\href{http://dx.doi.org/10.1016/0370-2693(82)90262-3}{{\em Phys.Lett.}
  {\bfseries B119} (1982) 136}.

\bibitem{Hall:1983id}
L.~J. Hall and M.~Suzuki
\href{http://dx.doi.org/10.1016/0550-3213(84)90513-3}{{\em Nucl.Phys.}
  {\bfseries B231} (1984) 419}.

\bibitem{Ellis:1984gi}
J.~R. Ellis, G.~Gelmini, C.~Jarlskog, G.~G. Ross, and J.~Valle
\href{http://dx.doi.org/10.1016/0370-2693(85)90157-1}{{\em Phys.Lett.}
  {\bfseries B150} (1985) 142}.

\bibitem{Ross:1984yg}
G.~G. Ross and J.~Valle
\href{http://dx.doi.org/10.1016/0370-2693(85)91658-2}{{\em Phys.Lett.}
  {\bfseries B151} (1985) 375}.

\bibitem{Dreiner:1997uz}
H.~K. Dreiner
\href{http://arxiv.org/abs/hep-ph/9707435}{{\ttfamily arXiv:hep-ph/9707435
  [hep-ph]}}.

\bibitem{Barbier:2004ez}
R.~Barbier, C.~Berat, M.~Besancon, M.~Chemtob, A.~Deandrea, {\em et~al.}
  \href{http://dx.doi.org/10.1016/j.physrep.2005.08.006}{{\em Phys.Rept.}
  {\bfseries 420} (2005) 1--202},
\href{http://arxiv.org/abs/hep-ph/0406039}{{\ttfamily arXiv:hep-ph/0406039
  [hep-ph]}}.

\bibitem{Csaki:2011ge}
C.~Csaki, Y.~Grossman, and B.~Heidenreich
  \href{http://dx.doi.org/10.1103/PhysRevD.85.095009}{{\em Phys.Rev.}
  {\bfseries D85} (2012) 095009},
\href{http://arxiv.org/abs/1111.1239}{{\ttfamily arXiv:1111.1239 [hep-ph]}}.

\bibitem{Csaki:2013we}
C.~Csaki and B.~Heidenreich
  \href{http://dx.doi.org/10.1103/PhysRevD.88.055023}{{\em Phys.Rev.}
  {\bfseries D88} (2013) 055023},
\href{http://arxiv.org/abs/1302.0004}{{\ttfamily arXiv:1302.0004 [hep-ph]}}.

\bibitem{Dimopoulos:1981yj}
S.~Dimopoulos, S.~Raby, and F.~Wilczek
\href{http://dx.doi.org/10.1103/PhysRevD.24.1681}{{\em Phys.Rev.} {\bfseries
  D24} (1981) 1681--1683}.

\bibitem{Raby:2011jt}
S.~Raby \href{http://dx.doi.org/10.1088/0034-4885/74/3/036901}{{\em
  Rept.Prog.Phys.} {\bfseries 74} (2011) 036901},
\href{http://arxiv.org/abs/1101.2457}{{\ttfamily arXiv:1101.2457 [hep-ph]}}.

\bibitem{Banks:1995by}
T.~Banks, Y.~Grossman, E.~Nardi, and Y.~Nir
  \href{http://dx.doi.org/10.1103/PhysRevD.52.5319}{{\em Phys.Rev.} {\bfseries
  D52} (1995) 5319--5325},
\href{http://arxiv.org/abs/hep-ph/9505248}{{\ttfamily arXiv:hep-ph/9505248
  [hep-ph]}}.

\bibitem{Smirnov:1995ey}
A.~Y. Smirnov and F.~Vissani
  \href{http://dx.doi.org/10.1016/0550-3213(95)00615-X}{{\em Nucl.Phys.}
  {\bfseries B460} (1996) 37--56},
\href{http://arxiv.org/abs/hep-ph/9506416}{{\ttfamily arXiv:hep-ph/9506416
  [hep-ph]}}.

\bibitem{Tamvakis:1996dk}
K.~Tamvakis \href{http://dx.doi.org/10.1016/0370-2693(96)00758-7}{{\em
  Phys.Lett.} {\bfseries B383} (1996) 307--312},
\href{http://arxiv.org/abs/hep-ph/9602389}{{\ttfamily arXiv:hep-ph/9602389
  [hep-ph]}}.

\bibitem{Barbieri:1997zn}
R.~Barbieri, A.~Strumia, and Z.~Berezhiani
  \href{http://dx.doi.org/10.1016/S0370-2693(97)00701-6}{{\em Phys.Lett.}
  {\bfseries B407} (1997) 250--254},
\href{http://arxiv.org/abs/hep-ph/9704275}{{\ttfamily arXiv:hep-ph/9704275
  [hep-ph]}}.

\bibitem{Giudice:1997wb}
G.~Giudice and R.~Rattazzi
  \href{http://dx.doi.org/10.1016/S0370-2693(97)00702-8}{{\em Phys.Lett.}
  {\bfseries B406} (1997) 321--327},
\href{http://arxiv.org/abs/hep-ph/9704339}{{\ttfamily arXiv:hep-ph/9704339
  [hep-ph]}}.

\bibitem{Tamvakis:1996np}
K.~Tamvakis \href{http://dx.doi.org/10.1016/0370-2693(96)00679-X}{{\em
  Phys.Lett.} {\bfseries B382} (1996) 251--256},
\href{http://arxiv.org/abs/hep-ph/9604343}{{\ttfamily arXiv:hep-ph/9604343
  [hep-ph]}}.

\bibitem{DiLuzio:2013ysa}
L.~Di~Luzio, M.~Nardecchia, and A.~Romanino
\href{http://arxiv.org/abs/1305.7034}{{\ttfamily arXiv:1305.7034 [hep-ph]}}.

\bibitem{Smirnov:1996bg}
A.~Y. Smirnov and F.~Vissani
  \href{http://dx.doi.org/10.1016/0370-2693(96)00495-9}{{\em Phys.Lett.}
  {\bfseries B380} (1996) 317--323},
\href{http://arxiv.org/abs/hep-ph/9601387}{{\ttfamily arXiv:hep-ph/9601387
  [hep-ph]}}.

\bibitem{Wells:2004di}
J.~D. Wells \href{http://dx.doi.org/10.1103/PhysRevD.71.015013}{{\em Phys.Rev.}
  {\bfseries D71} (2005) 015013},
\href{http://arxiv.org/abs/hep-ph/0411041}{{\ttfamily arXiv:hep-ph/0411041
  [hep-ph]}}.

\bibitem{ArkaniHamed:2004fb}
N.~Arkani-Hamed and S.~Dimopoulos
  \href{http://dx.doi.org/10.1088/1126-6708/2005/06/073}{{\em JHEP} {\bfseries
  0506} (2005) 073},
\href{http://arxiv.org/abs/hep-th/0405159}{{\ttfamily arXiv:hep-th/0405159
  [hep-th]}}.

\bibitem{Giudice:2004tc}
G.~Giudice and A.~Romanino
  \href{http://dx.doi.org/10.1016/j.nuclphysb.2004.11.048}{{\em Nucl.Phys.}
  {\bfseries B699} (2004) 65--89},
\href{http://arxiv.org/abs/hep-ph/0406088}{{\ttfamily arXiv:hep-ph/0406088
  [hep-ph]}}.

\bibitem{Gupta:2004dz}
S.~K. Gupta, P.~Konar, and B.~Mukhopadhyaya
  \href{http://dx.doi.org/10.1016/j.physletb.2004.12.014}{{\em Phys.Lett.}
  {\bfseries B606} (2005) 384--390},
\href{http://arxiv.org/abs/hep-ph/0408296}{{\ttfamily arXiv:hep-ph/0408296
  [hep-ph]}}.

\bibitem{Kuriyama:2008pv}
M.~Kuriyama, H.~Nakajima, and T.~Watari
  \href{http://dx.doi.org/10.1103/PhysRevD.79.075002}{{\em Phys.Rev.}
  {\bfseries D79} (2009) 075002},
\href{http://arxiv.org/abs/0802.2584}{{\ttfamily arXiv:0802.2584 [hep-ph]}}.

\bibitem{Mohapatra:1986aw}
R.~Mohapatra
\href{http://dx.doi.org/10.1103/PhysRevLett.56.561}{{\em Phys.Rev.Lett.}
  {\bfseries 56} (1986) 561--563}.

\bibitem{Nowakowski:1995dx}
M.~Nowakowski and A.~Pilaftsis
  \href{http://dx.doi.org/10.1016/0550-3213(95)00594-3}{{\em Nucl.Phys.}
  {\bfseries B461} (1996) 19--49},
\href{http://arxiv.org/abs/hep-ph/9508271}{{\ttfamily arXiv:hep-ph/9508271
  [hep-ph]}}.

\bibitem{Hempfling:1995wj}
R.~Hempfling \href{http://dx.doi.org/10.1016/0550-3213(96)00412-9}{{\em
  Nucl.Phys.} {\bfseries B478} (1996) 3--30},
\href{http://arxiv.org/abs/hep-ph/9511288}{{\ttfamily arXiv:hep-ph/9511288
  [hep-ph]}}.

\bibitem{deCarlos:1996du}
B.~de~Carlos and P.~White
  \href{http://dx.doi.org/10.1103/PhysRevD.54.3427}{{\em Phys.Rev.} {\bfseries
  D54} (1996) 3427--3446},
\href{http://arxiv.org/abs/hep-ph/9602381}{{\ttfamily arXiv:hep-ph/9602381
  [hep-ph]}}.

\bibitem{Nardi:1996iy}
E.~Nardi \href{http://dx.doi.org/10.1103/PhysRevD.55.5772}{{\em Phys.Rev.}
  {\bfseries D55} (1997) 5772--5779},
\href{http://arxiv.org/abs/hep-ph/9610540}{{\ttfamily arXiv:hep-ph/9610540
  [hep-ph]}}.

\bibitem{Chun:1998gp}
E.~Chun, S.~Kang, C.~Kim, and U.~Lee
  \href{http://dx.doi.org/10.1016/S0550-3213(99)00034-6}{{\em Nucl.Phys.}
  {\bfseries B544} (1999) 89--103},
\href{http://arxiv.org/abs/hep-ph/9807327}{{\ttfamily arXiv:hep-ph/9807327
  [hep-ph]}}.

\bibitem{Chun:1999bq}
E.~J. Chun and S.~K. Kang
  \href{http://dx.doi.org/10.1103/PhysRevD.61.075012}{{\em Phys.Rev.}
  {\bfseries D61} (2000) 075012},
\href{http://arxiv.org/abs/hep-ph/9909429}{{\ttfamily arXiv:hep-ph/9909429
  [hep-ph]}}.

\bibitem{Hirsch:2000ef}
M.~Hirsch, M.~Diaz, W.~Porod, J.~Romao, and J.~Valle
  \href{http://dx.doi.org/10.1103/PhysRevD.62.113008,
  10.1103/PhysRevD.65.119901}{{\em Phys.Rev.} {\bfseries D62} (2000) 113008},
\href{http://arxiv.org/abs/hep-ph/0004115}{{\ttfamily arXiv:hep-ph/0004115
  [hep-ph]}}.

\bibitem{Ade:2013zuv}
{\bfseries Planck Collaboration} Collaboration, P.~Ade {\em et~al.}
\href{http://arxiv.org/abs/1303.5076}{{\ttfamily arXiv:1303.5076
  [astro-ph.CO]}}.

\bibitem{Chikashige:1980qk}
Y.~Chikashige, R.~N. Mohapatra, and R.~Peccei
\href{http://dx.doi.org/10.1103/PhysRevLett.45.1926}{{\em Phys.Rev.Lett.}
  {\bfseries 45} (1980) 1926}.

\bibitem{Grossman:1998py}
Y.~Grossman and H.~E. Haber
  \href{http://dx.doi.org/10.1103/PhysRevD.59.093008}{{\em Phys.Rev.}
  {\bfseries D59} (1999) 093008},
\href{http://arxiv.org/abs/hep-ph/9810536}{{\ttfamily arXiv:hep-ph/9810536
  [hep-ph]}}.

\bibitem{Grossman:1999hc}
Y.~Grossman and H.~E. Haber
\href{http://arxiv.org/abs/hep-ph/9906310}{{\ttfamily arXiv:hep-ph/9906310
  [hep-ph]}}.

\bibitem{Davidson:2000uc}
S.~Davidson and M.~Losada {\em JHEP} {\bfseries 0005} (2000) 021,
\href{http://arxiv.org/abs/hep-ph/0005080}{{\ttfamily arXiv:hep-ph/0005080
  [hep-ph]}}.

\bibitem{Davidson:2000ne}
S.~Davidson and M.~Losada
  \href{http://dx.doi.org/10.1103/PhysRevD.65.075025}{{\em Phys.Rev.}
  {\bfseries D65} (2002) 075025},
\href{http://arxiv.org/abs/hep-ph/0010325}{{\ttfamily arXiv:hep-ph/0010325
  [hep-ph]}}.

\bibitem{Grossman:2003gq}
Y.~Grossman and S.~Rakshit
  \href{http://dx.doi.org/10.1103/PhysRevD.69.093002}{{\em Phys.Rev.}
  {\bfseries D69} (2004) 093002},
\href{http://arxiv.org/abs/hep-ph/0311310}{{\ttfamily arXiv:hep-ph/0311310
  [hep-ph]}}.

\bibitem{Nilles:1996ij}
H.-P. Nilles and N.~Polonsky
  \href{http://dx.doi.org/10.1016/S0550-3213(96)00511-1}{{\em Nucl.Phys.}
  {\bfseries B484} (1997) 33--62},
\href{http://arxiv.org/abs/hep-ph/9606388}{{\ttfamily arXiv:hep-ph/9606388
  [hep-ph]}}.

\bibitem{Martin:1993zk}
S.~P. Martin and M.~T. Vaughn \href{http://dx.doi.org/10.1103/PhysRevD.50.2282,
  10.1103/PhysRevD.78.039903}{{\em Phys.Rev.} {\bfseries D50} (1994) 2282},
\href{http://arxiv.org/abs/hep-ph/9311340}{{\ttfamily arXiv:hep-ph/9311340
  [hep-ph]}}.

\bibitem{Lee:1984kr}
I.-H. Lee
\href{http://dx.doi.org/10.1016/0370-2693(84)91885-9}{{\em Phys.Lett.}
  {\bfseries B138} (1984) 121}.

\bibitem{Lee:1984tn}
I.-H. Lee
\href{http://dx.doi.org/10.1016/0550-3213(84)90117-2}{{\em Nucl.Phys.}
  {\bfseries B246} (1984) 120}.

\bibitem{Littenberg:1991ek}
L.~S. Littenberg and R.~E. Shrock
\href{http://dx.doi.org/10.1103/PhysRevLett.68.443}{{\em Phys.Rev.Lett.}
  {\bfseries 68} (1992) 443--446}.

\bibitem{Littenberg:2000fg}
L.~S. Littenberg and R.~Shrock
  \href{http://dx.doi.org/10.1016/S0370-2693(00)01041-8}{{\em Phys.Lett.}
  {\bfseries B491} (2000) 285--290},
\href{http://arxiv.org/abs/hep-ph/0005285}{{\ttfamily arXiv:hep-ph/0005285
  [hep-ph]}}.

\bibitem{Leurer:1992wg}
M.~Leurer, Y.~Nir, and N.~Seiberg
  \href{http://dx.doi.org/10.1016/0550-3213(93)90112-3}{{\em Nucl.Phys.}
  {\bfseries B398} (1993) 319--342},
\href{http://arxiv.org/abs/hep-ph/9212278}{{\ttfamily arXiv:hep-ph/9212278
  [hep-ph]}}.

\bibitem{Dixon:1985jw}
L.~J. Dixon, J.~A. Harvey, C.~Vafa, and E.~Witten
\href{http://dx.doi.org/10.1016/0550-3213(85)90593-0}{{\em Nucl.Phys.}
  {\bfseries B261} (1985) 678--686}.

\bibitem{Dixon:1986jc}
L.~J. Dixon, J.~A. Harvey, C.~Vafa, and E.~Witten
\href{http://dx.doi.org/10.1016/0550-3213(86)90287-7}{{\em Nucl.Phys.}
  {\bfseries B274} (1986) 285--314}.

\bibitem{Ibanez:1986tp}
L.~E. Ibanez, H.~P. Nilles, and F.~Quevedo
\href{http://dx.doi.org/10.1016/0370-2693(87)90066-9}{{\em Phys.Lett.}
  {\bfseries B187} (1987) 25--32}.

\bibitem{Ibanez:1987sn}
L.~E. Ibanez, J.~E. Kim, H.~P. Nilles, and F.~Quevedo
\href{http://dx.doi.org/10.1016/0370-2693(87)90255-3}{{\em Phys.Lett.}
  {\bfseries B191} (1987) 282--286}.

\bibitem{Faraggi:2001ry}
A.~E. Faraggi \href{http://dx.doi.org/10.1016/S0370-2693(01)01165-0}{{\em
  Phys.Lett.} {\bfseries B520} (2001) 337--344},
\href{http://arxiv.org/abs/hep-ph/0107094}{{\ttfamily arXiv:hep-ph/0107094
  [hep-ph]}}.

\bibitem{Witten:1985xc}
E.~Witten
\href{http://dx.doi.org/10.1016/0550-3213(85)90603-0}{{\em Nucl.Phys.}
  {\bfseries B258} (1985) 75}.

\bibitem{Lebedev:2006kn}
O.~Lebedev, H.~P. Nilles, S.~Raby, S.~Ramos-Sanchez, M.~Ratz, {\em et~al.}
  \href{http://dx.doi.org/10.1016/j.physletb.2006.12.012}{{\em Phys.Lett.}
  {\bfseries B645} (2007) 88--94},
\href{http://arxiv.org/abs/hep-th/0611095}{{\ttfamily arXiv:hep-th/0611095
  [hep-th]}}.

\bibitem{Lebedev:2007hv}
O.~Lebedev, H.~P. Nilles, S.~Raby, S.~Ramos-Sanchez, M.~Ratz, {\em et~al.}
  \href{http://dx.doi.org/10.1103/PhysRevD.77.046013}{{\em Phys.Rev.}
  {\bfseries D77} (2008) 046013},
\href{http://arxiv.org/abs/0708.2691}{{\ttfamily arXiv:0708.2691 [hep-th]}}.

\bibitem{Lebedev:2008un}
O.~Lebedev, H.~P. Nilles, S.~Ramos-Sanchez, M.~Ratz, and P.~K. Vaudrevange
  \href{http://dx.doi.org/10.1016/j.physletb.2008.08.054}{{\em Phys.Lett.}
  {\bfseries B668} (2008) 331--335},
\href{http://arxiv.org/abs/0807.4384}{{\ttfamily arXiv:0807.4384 [hep-th]}}.

\bibitem{Blaszczyk:2009in}
M.~Blaszczyk, S.~Groot~Nibbelink, M.~Ratz, F.~Ruehle, M.~Trapletti, {\em
  et~al.} \href{http://dx.doi.org/10.1016/j.physletb.2009.12.036}{{\em
  Phys.Lett.} {\bfseries B683} (2010) 340--348},
\href{http://arxiv.org/abs/0911.4905}{{\ttfamily arXiv:0911.4905 [hep-th]}}.

\bibitem{Kawamura:2000ev}
Y.~Kawamura \href{http://dx.doi.org/10.1143/PTP.105.999}{{\em Prog.Theor.Phys.}
  {\bfseries 105} (2001) 999--1006},
\href{http://arxiv.org/abs/hep-ph/0012125}{{\ttfamily arXiv:hep-ph/0012125
  [hep-ph]}}.

\bibitem{Hall:2001pg}
L.~J. Hall and Y.~Nomura
  \href{http://dx.doi.org/10.1103/PhysRevD.64.055003}{{\em Phys.Rev.}
  {\bfseries D64} (2001) 055003},
\href{http://arxiv.org/abs/hep-ph/0103125}{{\ttfamily arXiv:hep-ph/0103125
  [hep-ph]}}.

\bibitem{Hebecker:2001wq}
A.~Hebecker and J.~March-Russell
  \href{http://dx.doi.org/10.1016/S0550-3213(01)00374-1}{{\em Nucl.Phys.}
  {\bfseries B613} (2001) 3--16},
\href{http://arxiv.org/abs/hep-ph/0106166}{{\ttfamily arXiv:hep-ph/0106166
  [hep-ph]}}.

\bibitem{th/9506150}
G.~Papadopoulos and P.~Townsend
  \href{http://dx.doi.org/10.1016/0370-2693(95)00929-F}{{\em Phys.Lett.}
  {\bfseries B357} (1995) 300--306},
\href{http://arxiv.org/abs/hep-th/9506150}{{\ttfamily arXiv:hep-th/9506150
  [hep-th]}}.

\bibitem{th/0011089}
B.~S. Acharya
\href{http://arxiv.org/abs/hep-th/0011089}{{\ttfamily arXiv:hep-th/0011089
  [hep-th]}}.

\bibitem{th/0107177}
M.~Atiyah and E.~Witten {\em Adv.Theor.Math.Phys.} {\bfseries 6} (2003) 1--106,
\href{http://arxiv.org/abs/hep-th/0107177}{{\ttfamily arXiv:hep-th/0107177
  [hep-th]}}.

\bibitem{th/0109152}
B.~S. Acharya and E.~Witten
\href{http://arxiv.org/abs/hep-th/0109152}{{\ttfamily arXiv:hep-th/0109152
  [hep-th]}}.

\bibitem{th/0701034}
B.~S. Acharya, K.~Bobkov, G.~L. Kane, P.~Kumar, and J.~Shao
  \href{http://dx.doi.org/10.1103/PhysRevD.76.126010}{{\em Phys.Rev.}
  {\bfseries D76} (2007) 126010},
\href{http://arxiv.org/abs/hep-th/0701034}{{\ttfamily arXiv:hep-th/0701034
  [hep-th]}}.

\bibitem{0801.0478}
B.~S. Acharya, K.~Bobkov, G.~L. Kane, J.~Shao, and P.~Kumar
  \href{http://dx.doi.org/10.1103/PhysRevD.78.065038}{{\em Phys.Rev.}
  {\bfseries D78} (2008) 065038},
\href{http://arxiv.org/abs/0801.0478}{{\ttfamily arXiv:0801.0478 [hep-ph]}}.

\bibitem{ph/0201018}
E.~Witten
\href{http://arxiv.org/abs/hep-ph/0201018}{{\ttfamily arXiv:hep-ph/0201018
  [hep-ph]}}.

\bibitem{Csaki:2013jza}
C.~Csaki, E.~Kuflik, and T.~Volansky
\href{http://arxiv.org/abs/1309.5957}{{\ttfamily arXiv:1309.5957 [hep-ph]}}.

\bibitem{Dvali:1996cu}
G.~Dvali, G.~Giudice, and A.~Pomarol
  \href{http://dx.doi.org/10.1016/0550-3213(96)00404-X}{{\em Nucl.Phys.}
  {\bfseries B478} (1996) 31--45},
\href{http://arxiv.org/abs/hep-ph/9603238}{{\ttfamily arXiv:hep-ph/9603238
  [hep-ph]}}.

\bibitem{Randall:1998uk}
L.~Randall and R.~Sundrum
  \href{http://dx.doi.org/10.1016/S0550-3213(99)00359-4}{{\em Nucl.Phys.}
  {\bfseries B557} (1999) 79--118},
\href{http://arxiv.org/abs/hep-th/9810155}{{\ttfamily arXiv:hep-th/9810155
  [hep-th]}}.

\bibitem{Kim:1983dt}
J.~E. Kim and H.~P. Nilles
\href{http://dx.doi.org/10.1016/0370-2693(84)91890-2}{{\em Phys.Lett.}
  {\bfseries B138} (1984) 150}.

\bibitem{Casas:1992mk}
J.~Casas and C.~Munoz
  \href{http://dx.doi.org/10.1016/0370-2693(93)90081-R}{{\em Phys.Lett.}
  {\bfseries B306} (1993) 288--294},
\href{http://arxiv.org/abs/hep-ph/9302227}{{\ttfamily arXiv:hep-ph/9302227
  [hep-ph]}}.

\bibitem{Giudice:1988yz}
G.~Giudice and A.~Masiero
\href{http://dx.doi.org/10.1016/0370-2693(88)91613-9}{{\em Phys.Lett.}
  {\bfseries B206} (1988) 480--484}.

\bibitem{1009.0905}
H.~M. Lee, S.~Raby, M.~Ratz, G.~G. Ross, R.~Schieren, {\em et~al.}
  \href{http://dx.doi.org/10.1016/j.physletb.2010.10.038}{{\em Phys.Lett.}
  {\bfseries B694} (2011) 491--495},
\href{http://arxiv.org/abs/1009.0905}{{\ttfamily arXiv:1009.0905 [hep-ph]}}.

\bibitem{1102.3595}
H.~M. Lee, S.~Raby, M.~Ratz, G.~G. Ross, R.~Schieren, {\em et~al.}
  \href{http://dx.doi.org/10.1016/j.nuclphysb.2011.04.009}{{\em Nucl.Phys.}
  {\bfseries B850} (2011) 1--30},
\href{http://arxiv.org/abs/1102.3595}{{\ttfamily arXiv:1102.3595 [hep-ph]}}.

\bibitem{Lebedev:2006tr}
O.~Lebedev, H.-P. Nilles, S.~Raby, S.~Ramos-Sanchez, M.~Ratz, {\em et~al.}
  \href{http://dx.doi.org/10.1103/PhysRevLett.98.181602}{{\em Phys.Rev.Lett.}
  {\bfseries 98} (2007) 181602},
\href{http://arxiv.org/abs/hep-th/0611203}{{\ttfamily arXiv:hep-th/0611203
  [hep-th]}}.

\bibitem{Kappl:2008ie}
R.~Kappl, H.~P. Nilles, S.~Ramos-Sanchez, M.~Ratz, K.~Schmidt-Hoberg, {\em
  et~al.} \href{http://dx.doi.org/10.1103/PhysRevLett.102.121602}{{\em
  Phys.Rev.Lett.} {\bfseries 102} (2009) 121602},
\href{http://arxiv.org/abs/0812.2120}{{\ttfamily arXiv:0812.2120 [hep-th]}}.

\bibitem{Brummer:2010fr}
F.~Brummer, R.~Kappl, M.~Ratz, and K.~Schmidt-Hoberg
  \href{http://dx.doi.org/10.1007/JHEP04(2010)006}{{\em JHEP} {\bfseries 1004}
  (2010) 006},
\href{http://arxiv.org/abs/1003.0084}{{\ttfamily arXiv:1003.0084 [hep-th]}}.

\bibitem{Affleck:1983mk}
I.~Affleck, M.~Dine, and N.~Seiberg
\href{http://dx.doi.org/10.1016/0550-3213(84)90058-0}{{\em Nucl.Phys.}
  {\bfseries B241} (1984) 493--534}.

\bibitem{Chen:2012jg}
M.-C. Chen, M.~Ratz, C.~Staudt, and P.~K. Vaudrevange
  \href{http://dx.doi.org/10.1016/j.nuclphysb.2012.08.018}{{\em Nucl.Phys.}
  {\bfseries B866} (2013) 157--176},
\href{http://arxiv.org/abs/1206.5375}{{\ttfamily arXiv:1206.5375 [hep-ph]}}.

\bibitem{Cleaver:1997jb}
G.~Cleaver, M.~Cvetic, J.~R. Espinosa, L.~L. Everett, and P.~Langacker
  \href{http://dx.doi.org/10.1016/S0550-3213(98)00277-6}{{\em Nucl.Phys.}
  {\bfseries B525} (1998) 3--26},
\href{http://arxiv.org/abs/hep-th/9711178}{{\ttfamily arXiv:hep-th/9711178
  [hep-th]}}.

\bibitem{Dine:1987xk}
M.~Dine, N.~Seiberg, and E.~Witten
\href{http://dx.doi.org/10.1016/0550-3213(87)90395-6}{{\em Nucl.Phys.}
  {\bfseries B289} (1987) 589}.

\bibitem{Atick:1987gy}
J.~J. Atick, L.~J. Dixon, and A.~Sen
\href{http://dx.doi.org/10.1016/0550-3213(87)90639-0}{{\em Nucl.Phys.}
  {\bfseries B292} (1987) 109--149}.

\bibitem{1102.0556}
B.~S. Acharya, G.~Kane, E.~Kuflik, and R.~Lu
  \href{http://dx.doi.org/10.1007/JHEP05(2011)033}{{\em JHEP} {\bfseries 1105}
  (2011) 033},
\href{http://arxiv.org/abs/1102.0556}{{\ttfamily arXiv:1102.0556 [hep-ph]}}.

\bibitem{th/0203061}
C.~Beasley and E.~Witten {\em JHEP} {\bfseries 0207} (2002) 046,
\href{http://arxiv.org/abs/hep-th/0203061}{{\ttfamily arXiv:hep-th/0203061
  [hep-th]}}.

\bibitem{Choi:1997de}
K.~Choi, J.~S. Lee, and C.~Munoz
  \href{http://dx.doi.org/10.1103/PhysRevLett.80.3686}{{\em Phys.Rev.Lett.}
  {\bfseries 80} (1998) 3686--3689},
\href{http://arxiv.org/abs/hep-ph/9709250}{{\ttfamily arXiv:hep-ph/9709250
  [hep-ph]}}.

\bibitem{Kuflik:2010dg}
E.~Kuflik and J.~Marsano \href{http://dx.doi.org/10.1007/JHEP03(2011)020}{{\em
  JHEP} {\bfseries 1103} (2011) 020},
\href{http://arxiv.org/abs/1009.2510}{{\ttfamily arXiv:1009.2510 [hep-ph]}}.

\bibitem{Martin:1992mq}
S.~P. Martin \href{http://dx.doi.org/10.1103/PhysRevD.46.R2769}{{\em Phys.Rev.}
  {\bfseries D46} (1992) 2769--2772},
\href{http://arxiv.org/abs/hep-ph/9207218}{{\ttfamily arXiv:hep-ph/9207218
  [hep-ph]}}.

\bibitem{FileviezPerez:2008sx}
P.~Fileviez~Perez and S.~Spinner
  \href{http://dx.doi.org/10.1016/j.physletb.2009.02.047}{{\em Phys.Lett.}
  {\bfseries B673} (2009) 251--254},
\href{http://arxiv.org/abs/0811.3424}{{\ttfamily arXiv:0811.3424 [hep-ph]}}.

\bibitem{Barger:2008wn}
V.~Barger, P.~Fileviez~Perez, and S.~Spinner
  \href{http://dx.doi.org/10.1103/PhysRevLett.102.181802}{{\em Phys.Rev.Lett.}
  {\bfseries 102} (2009) 181802},
\href{http://arxiv.org/abs/0812.3661}{{\ttfamily arXiv:0812.3661 [hep-ph]}}.

\bibitem{Ambroso:2009jd}
M.~Ambroso and B.~Ovrut
  \href{http://dx.doi.org/10.1088/1126-6708/2009/10/011}{{\em JHEP} {\bfseries
  0910} (2009) 011},
\href{http://arxiv.org/abs/0904.4509}{{\ttfamily arXiv:0904.4509 [hep-th]}}.

\bibitem{Ambroso:2010pe}
M.~Ambroso and B.~A. Ovrut
  \href{http://dx.doi.org/10.1142/S0217751X11052943}{{\em Int.J.Mod.Phys.}
  {\bfseries A26} (2011) 1569--1627},
\href{http://arxiv.org/abs/1005.5392}{{\ttfamily arXiv:1005.5392 [hep-th]}}.

\bibitem{Marshall:2014kea}
Z.~Marshall, B.~A. Ovrut, A.~Purves, and S.~Spinner
  \href{http://dx.doi.org/10.1016/j.physletb.2014.03.052}{{\em Phys.Lett.}
  {\bfseries B732} (2014) 325--329},
\href{http://arxiv.org/abs/1401.7989}{{\ttfamily arXiv:1401.7989 [hep-ph]}}.

\bibitem{Braun:2005nv}
V.~Braun, Y.-H. He, B.~A. Ovrut, and T.~Pantev
  \href{http://dx.doi.org/10.1088/1126-6708/2006/05/043}{{\em JHEP} {\bfseries
  0605} (2006) 043},
\href{http://arxiv.org/abs/hep-th/0512177}{{\ttfamily arXiv:hep-th/0512177
  [hep-th]}}.

\end{thebibliography}\endgroup

\end{document}